\documentclass[12pt, letter]{article}

\usepackage{amsmath}
\usepackage{amssymb}
\usepackage{array}
\usepackage{amsthm}
\usepackage{amsbsy}
\usepackage{amsfonts}
\usepackage{dsfont}
\usepackage{color}
\usepackage{graphicx}
\usepackage[dvips]{epsfig}
\usepackage{harvard}
\usepackage[latin1]{inputenc}

\renewcommand{\baselinestretch}{1.2}
\usepackage[margin=1in]{geometry}
\graphicspath{{figs/}}
\DeclareGraphicsExtensions{.eps.gz,.eps,.epsi.gz,.epsi,.ps,.ps.gz}
\setkeys{Gin}{width=\textwidth}
\newcommand{\bm}[1]{{\boldsymbol {#1}}}

\newcommand{\eps}{\epsilon}

\title{Approximating Probability Densities by Iterated Laplace
  Approximations}

\author{Björn Bornkamp\footnote{Work done while Björn Bornkamp was
    Postdoctoral Fellow at Fakultät Statistik, Technische Universität
    Dortmund, 44221 Dortmund, Germany (email:
    bornkamp@statistik.tu-dortmund.de).}}

\begin{document}
\maketitle \vspace*{-1.2cm}

\thispagestyle{empty}
\begin{abstract}
  The Laplace approximation is an old, but frequently used method to
  approximate integrals for Bayesian calculations. In this paper we
  develop an extension of the Laplace approximation, by applying it
  iteratively to the residual, \textit{i.e.}, the difference between
  the current approximation and the true function. The final
  approximation is thus a linear combination of multivariate normal
  densities, where the coefficients are chosen to achieve a good fit
  to the target distribution. We illustrate on real and artificial
  examples that the proposed procedure is a computationally efficient
  alternative to current approaches for approximation of multivariate
  probability densities.
\end{abstract}
\textbf{keywords:} Bayes Factor, Importance Sampling, Markov Chain
Monte Carlo, Non-Linear Regression, Normalization Constant

\newpage \setcounter{page}{1}
\section{Introduction}
\label{sec:intro}

Suppose you are given a positive integrable function $\pi(\bm x)$ on
$\mathbb{R}^p$, with unknown normalization constant $Z=\int \pi(\bm
x)\mathrm{d}\bm x$ and it is desired to approximate the probability
distribution $\pi(\bm x)/Z$. In the context of Bayesian statistics
this could for example be the posterior distribution. The idea of the
standard Laplace approximation is to maximize $\log(\pi(\bm x))$,
resulting in the mode $\tilde{\bm x}$ (\textit{i.e.} a point with
gradient 0) and to approximate $\log(\pi(\bm x))$ by a second order
Taylor approximation in $\tilde{\bm x}$ (assuming the necessary
derivatives of $\log(\pi(\bm x))$ exist). This leads to an
approximation of the form
\begin{equation*}
  \pi(\bm x) \approx  \exp(\log(\pi(\tilde{\bm x})) + 1/2(\bm x
  - \tilde{\bm x})'H(\tilde{\bm x})(\bm x - \tilde{\bm x})),
\end{equation*}
where $H(\tilde{\bm x}) = \partial \log{\pi({\bm x})}/\partial
x_i \partial x_j$ is the Hesse matrix of $\log({\pi({\bm x})})$
evaluated at $\tilde{\bm x}$. Essentially, $\pi(\bm x)$ is
approximated by the kernel of a multivariate normal distribution with
mean $\tilde{\bm x}$ and covariance matrix $\tilde{\bm
  \Sigma}=-H(\tilde{\bm x})^{-1}$. The normalization constant of this
approximation is $(2\pi)^{p/2}|\tilde{\bm \Sigma}|^{1/2}\pi(\tilde{\bm
  x})$, which itself approximates $Z$.  At first sight it might appear
simplistic to approximate any distribution by a normal distribution,
but in the context of Bayesian statistics this approach is justified
by the asymptotic normality of posterior distributions, and often
works also for moderate sample sizes; see, for example,
\citeasnoun[Chapter 4]{evan:swar:2000} or \citeasnoun{tier:kada:1986}
for more details and \citeasnoun{dici:kass:raft:1997} or
\citeasnoun{nott:kohn:fiel:2008} for further work in the context of
approximating the normalization constant. The recent article by
\citeasnoun{rue:mart:chop:2009} successfully uses Laplace
approximations for latent variables in latent Gaussian models to
approximate marginal densities, while \citeasnoun{hara:tier:2010} use
the Laplace approximation to build automatic MCMC algorithms in a
related model class.

The Laplace approximation is always unimodal and elliptical. In the
case of multiple modes, one can partially overcome this problem by
fitting Laplace approximations to each mode (see \citeasnoun[Chapter
12]{gelm:carl:ster:2003}). Nevertheless, non-elliptical skew posterior
distributions remain a challenge and relatively few papers try to
improve the Laplace approximation in this regard. One approach is to
employ third order derivatives in the Taylor expansion (see for
example \citeasnoun[p. 238--239]{ohag:fors:2004}). Unfortunately,
these can be hard to calculate, particularly in larger dimensions.
Due to dominance of the cubic terms, the approximation might also
diverge in the tails, so that it might not be integrable. Another
approach is to center the normal approximation on the first two
moments, instead of mode and negative inverse Hessian at the mode
\cite{mink:2001}.  \citeasnoun{nott:fiel:leon:2009} consider to
improve the Laplace approximation based on numerical integration, in
particular for approximating $Z$.

In this paper the idea of iterating the Laplace approximation is
developed. Given an approximation $\tilde{\pi}(\bm x)$ of $\pi(\bm
x)$, the approximation of $\pi(\bm x)$ is improved by fitting a
Laplace approximation to the residual $r(\bm x)=\pi(\bm
x)-\tilde{\pi}(\bm x)$. The current approximation $\tilde{\pi}(\bm x)$
fits worst where $r(\bm x)$ has its maximum, so that the resulting
residual Laplace approximation corrects the deficiencies of fit
between $\tilde{\pi}(\bm x)$ and $\pi(\bm x)$. The new approximation
is then given by a linear combination of the starting approximation
$\tilde{\pi}(\bm x)$ and the residual Laplace approximation. This
process is repeated until the approximation does not change
considerably. The idea of improving an approximation by fitting the
``residuals'' of the current approximation and then using a linear
combination of the obtained functions, has been applied elsewhere in
the statistical literature. It is, for example, at the core of the
relaxed greedy algorithm described in \citeasnoun{barr:cohe:dahm:2008}
(who use dictionaries to model the individual functions) or the
boosting technique in machine learning, where this idea is usually
called functional gradient descent, see \citeasnoun{bueh:yu:2003}. The
difference to the procedure presented here is that the Laplace
approximation is used to define new elements in the linear
combination, rather than relying on a functional basis or
dictionaries. Using Laplace approximation has the advantage that it is
automatically centered at the point, where the current approximation
is worst, \textit{i.e.}, where the residual is largest. In addition,
the approximation is a linear combination of multivariate normal
densities. This is convenient from a statistical and computational
viewpoint (\textit{e.g.},  sampling random variates and evaluation of
the density are straightforward).

At the end of the iterative process one hence obtains a global
approximation of $\pi(\bm x)$, which can be used for subsequent
statistical inference directly, or as a proposal distribution for
Monte Carlo simulation (such as importance sampling or the
independence Metropolis-Hastings algorithm
\cite[ch. 2.3]{tier:1994}). Building global approximations to obtain
proposal distributions has received considerable attention in recent
years. An idea appearing repeatedly is that of building a mixture
based approximation, while performing the iterative simulation
algorithm. In the context of importance sampling this has been
considered for example by \citeasnoun{capp:douc:guil:2008} or
\citeasnoun{ardi:hoog:vand:2009}. In the context of adaptive MCMC,
mixture based approximations have been discussed for example by
\citeasnoun[ch. 7]{andr:moul:2006} or \citeasnoun{gior:kohn:2010}.

We think that iterating Laplace approximations can be a viable
alternative to current approaches for building global approximations
of $\pi(\bm x)/Z$.  In Section \ref{sec:laplappr} this idea will be
elaborated in detail, while in Section \ref{sec:appl} the methodology
will be evaluated on three test examples and a nonlinear regression
application.

\section{Iterated Laplace Approximations}
\label{sec:laplappr}

The main idea of iterated Laplace approximations, abbreviated
\textsf{iterLap}, is to improve the current approximation
$\tilde{\pi}(\bm x)$ of $\pi(\bm x)$ by performing a Laplace
approximation of the residual $r(\bm x) = \pi(\bm x) - \tilde{\pi}(\bm
x)$, and then taking $w_1\tilde{\pi}(\bm x)+w_2r_L(\bm x)$ as the new
approximation, where $r_L(\bm x)$ is the (normalized) Laplace
approximation of $r(\bm x)$ (\textit{i.e.}, a multivariate normal
density) and the $w_j$ are suitably determined positive
coefficients. This procedure is then iterated until a satisfying
approximation is achieved. The final approximation of $\pi(\bm x)$
will be a linear combination of multivariate normal densities and the
normalizing constant of the approximation can be obtained simply by
adding the coefficients $w_j$. The residual $r(\bm x)$ can get
negative, which causes a problem when calculating $\log(r(\bm
x))$. Hence one should use, for example, the positive function $r(\bm
x)1_{A}(\bm x)+\exp(r(\bm x)-\tilde{\eps})\tilde{\eps}1_{A^c}(\bm x)$
as the objective function, where $A = \{x|r(\bm x)\geq \tilde{\eps}\}$
and $\tilde{\eps}$ is a small positive constant. This does not change
the residual in the region of interest, where $r(\bm x)>0$.

\renewcommand{\baselinestretch}{1.25}
\begin{figure}
\centering
\centerline{\includegraphics[width=0.9\textwidth]{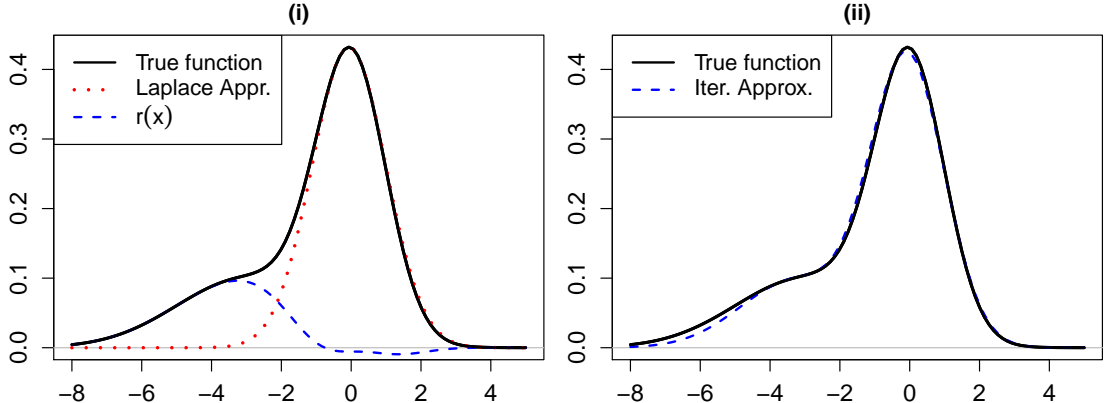}}
\caption{Illustrating the iterated Laplace approximation.}
  \label{fig:illustration}
\end{figure}
\renewcommand{\baselinestretch}{2}

The procedure will be illustrated here by fitting the one-dimensional
skew non-normalized density $\pi(x)=\phi(x,0,1)+0.5\phi(x,-3,2)$,
where $\phi(x,\mu,\sigma)$ denotes the density of the normal
distribution with mean $\mu$ and standard deviation $\sigma$. In
Figure \ref{fig:illustration} (i) the true function $\pi(x)$ and the
Laplace approximation $\tilde{\pi}(x)$ based on the global maximum are
displayed. Due to skewness, the Laplace approximation fits poorly in
the left part of $\pi(x)$, where the residual function $r(x)$
consequently has its maximum. Figure \ref{fig:illustration} (ii)
displays the new approximation based on a mixture of the initial
Laplace approximation and the Laplace approximation of the residual
$r(x)$. The coefficients $w_1$ and $w_2$ of the two components were
chosen to minimize the $L_2$ distance between approximation and truth
$\pi(x)$.  The approximation fits the true function fairly well,
despite a remaining small discrepancy in the left tail, which could be
eliminated by a further iteration of the algorithm.

While the conceptual idea of iterating Laplace approximations is
fairly simple, care must be taken when implementing the idea.  In the
following the computational details of iterated Laplace approximations
will be described.

\textbf{Algorithm}\\
\textit{Iteration 0:}
\begin{enumerate}
\item Fit a Laplace approximation to each mode of $\pi(\bm x)$, to
  obtain a starting approximation: $\tilde{\pi}_0(\bm
  x)=\sum_{j=1}^{J^{(0)}}w_j\phi(\bm x, \bm \mu_j, \bm \Sigma_j)$,
  where $\phi(\bm x, \bm \mu, \bm \Sigma)$ denotes the density of a
  multivariate normal distribution with mean $\bm \mu$ and covariance
  matrix $\bm \Sigma$, $J^{(0)}$ the number of found modes, $\bm
  \mu_j$ the modes, $\bm \Sigma_j$ the negative inverse Hessians of
  $\log(\pi(\bm x))$ evaluated at the modes and $w_j =(2\pi)^{p/2}|\bm
  \Sigma_j|^{1/2}\pi(\bm \mu_j)$; see \citeasnoun[Chapter
  12]{gelm:carl:ster:2003} for details on this multiple mode Laplace
  approximation.
\item Determine for each component in the linear combination a grid of
  size $n$ that encloses most of its probability mass (see A.1 below
  for more details). Let $\bm X_0$ denote the $nJ^{(0)} \times p$
  matrix that contains these grid points in the rows.
\item Evaluate $\pi(.)$ at $\bm X_0$, resulting in the vector $\bm
  y_0$ of length $nJ^{(0)}$. Also evaluate each of the $J^{(0)}$
  component densities in the mixture at $\bm X_0$ and write those
  evaluations in the $nJ^{(0)} \times J^{(0)}$ matrix $\bm F_0$.
\item Check the stopping criterion (see A.2 below for details); if
  this is not met initialize $t \leftarrow 1$.
\end{enumerate}

\textit{Iteration t:}
\begin{enumerate}
\item In this step the residual Laplace approximation is performed to
  obtain one new mixture component. Select $k$ possible starting
  values for optimization of the log-residual $\log(r(\bm x))$, with
  $r(\bm x)=\pi(\bm x)-\tilde{\pi}_t(\bm x)$ (see A.3 below for
  details on selecting starting values). Start a local optimizer at
  the first starting value, potentially resulting in a maximum
  $\tilde{\bm x}$ with zero gradient vector. If the Hesse matrix
  $\tilde{\bm H}$ at $\tilde{\bm x}$ is negative definite, a new
  mixture component has been found, \textit{i.e.}, one increments the
  number of components $J^{(t)}\leftarrow J^{(t-1)}+1$, and sets $\bm
  \mu_{J^{(t)}}=\tilde{\bm x}$ and $\bm \Sigma_{J^{(t)}} = -\tilde{\bm
    H}^{-1}$. Otherwise, if $\tilde{\bm H}$ is not negative definite
  one tries the next starting value. If all of the $k$ starting values
  fail, stop the procedure, as no adequate improvement can be found;
  otherwise continue to step 2.
\item Determine a grid $\bm N_t$ of size $n$ for the new component
  that encloses most of its probability mass (see A.1). Add these
  points to the current grid $\bm X_{t-1}$ to form $\bm X_{t}=
  \left( \begin{smallmatrix} \bm X_{t-1} \\ \bm
      N_t \end{smallmatrix}\right)$, which is then of size
  $nJ^{(t)}\times p$.
\item Evaluate $\pi(.)$ at $\bm N_t$ and append these evaluations to
  $\bm y_{t-1}$ to form $\bm y_t$. Evaluate all components of the
  approximation $\tilde{\pi}_{t-1}(.)$ at $\bm N_t$ and the new
  component at the entire grid $\bm X_t$ to form the $nJ^{(t)} \times
  J^{(t)}$ matrix $\bm F_t$.
\item Find the coefficients $w_1,\ldots,w_{J^{(t)}}$ by minimizing
  $(\bm y_{t}-\bm F_{t}'\bm w)'(\bm y_{t}-\bm F_{t}'\bm w)$ subject to
  $w_j \geq 0$ for $j=1,\ldots,J^{(t)}$ (see A.4 below for
  details) and calculate the current approximation of the
  normalization constant $Z_t=\sum_{j=1}^{J^{(t)}}w_j$. The current
  approximation of $\pi(\bm x)$ is then $\tilde{\pi}_t(\bm x) =
  \sum_{j=1}^{J^{(t)}}w_j\phi(\bm x, \bm \mu_j,\bm \Sigma_j)$.
\item Check the stopping criteria, if they are not met iterate $t
  \leftarrow t+1$ (see also A.2).
\end{enumerate}

The points below illustrate computational details in the
implementation of the algorithm.
\begin{itemize}
\item[A.1] \textbf{Grid}\\
  The reason for choosing the grid is to identify regions where
  $\pi(.)$ has positive probability mass. In our experience it works
  well to use a quasi-random sample of the multivariate normal
  distribution underlying the selected component. For this purpose a
  randomized quasi-random sample generated by the Sobol sequence is
  used (as implemented for example in the R package
  \texttt{randtoolbox}, \citeasnoun{duta:2009}). Compared to a pseudo
  random sample this has the advantage that the space is more
  systematically covered.  A default choice of $n$ is discussed below.
\item[A.2] \textbf{Stopping Criteria}\\
  Different criteria can be used for stopping the iterative process.
  First one can compare $\bm y_t$ and $\tilde{\bm y}_t=\bm F_t'\bm
  w_t$, \textit{i.e.}  stopping the iterative process, when
  ${\max}|\bm y_t-\tilde{\bm y}_t|< \delta M_t,$ where
  $M_t=\underset{\bm X_t}{\max}\, \pi(\bm x)$ and $\delta$ is a small
  positive number. This criterion assesses the quality of the
  approximation on the current grid, and stops the process, when there
  are only small differences between truth and approximation.  This
  criterion is already available at iteration 0. Another stopping rule
  is to monitor the normalizing constant $Z_t$ of $\tilde{\pi}_t(\bm
  x)$. It measures the ``volume'' of the approximation
  $\tilde{\pi}_t(\bm x)$; if $Z_t$ does not change further this
  indicates that the algorithm cannot find more regions, where
  $\pi(\bm x)$ has relevant probability mass. It is not uncommon that
  two consecutive iterations only lead to small changes in the
  normalizing constant, so we stop the iterative process, when $Z_t$
  does not change considerably the third time in a row, \textit{i.e.},
  when $\frac{|Z_t-0.5(Z_{t-1}+Z_{t-2})|}{Z_t}<\eps$, where $\eps$ is
  a small positive constant. Note that both of the above stopping
  criteria (as most stopping criteria for iterative calculations), do
  not guarantee a certain quality of the solution, but with sensible
  values for $\eps$ and $\delta$ (default choices are discussed below)
  one often obtains a satisfactory result. Note that the procedure can
  also stop, when no adequate modes can be found in step 1 of
  iteration $t$, or when a maximal pre-specified number of components
  $T$ is reached. All of the above criteria will be employed in the
  examples discussed in the later sections.
\item[A.3] \textbf{Starting values}\\
  The residual function $r(\bm x)$ is often multimodal. Because the
  used optimizers are designed for finding local optima, it is crucial
  to use good starting values. In our experience it works well to use
  starting values where $y_{i}/\tilde{y}_{i},\;i=1,\ldots,nJ^{(t)}$ is
  largest; here $y_i$ and $\tilde{y}_{i}$ denote the entries of $\bm
  y_t$ and $\tilde{\bm y}_t$. This choice works better than, for
  example, using the values where $y_i-\tilde{y}_{i}$ is largest,
  because the former selects values further apart from the current
  modes. In our implementation the ten grid values for which the
  $y_{i}/\tilde{y}_{i}$ is largest are selected, and then clustered
  with the $k$-means algorithm \cite{hart:wong:1979} to obtain $k=3$
  starting values. The order in which the starting values are used is
  determined by the distance to the last added mode, with the values
  farthest away being tried first. This prevents the algorithm from
  wandering in only one direction.
\item[A.4] \textbf{Quadratic Programming}\\
  Solving the constrained least squares problem in step 4 is a
  quadratic programming problem and can be solved efficiently for
  example with the algorithm of
  \citeasnoun{gold:idna:1982}.
\end{itemize}

In summary the algorithm needs: The grid size $n$ for each component,
the values $\eps$ and $\delta$ for the two stopping criteria and the
maximum number $T$ of components allowed. Suited default values for
those parameters have been determined by experimentation based on a
number of example densities, covering a range of different cases
observed in practice. A good time-quality trade off for the grid size
was obtained for the smallest integer larger than $50p^{1.25}$. The
number $50$ has been selected based on the observation that a grid
size of $50$ is often sufficient in one dimensional cases, the
exponent $1.25$ has been chosen to obtain a grid size that grows
slightly faster than linear in the dimension. For $\delta$ and $\eps$
the choices $\delta=0.01$ and $\eps = 0.005$ were found to work
well. The main rationale for these selections is that one should not
stop the process before all relevant probability mass has been
identified. On the other hand, stopping late will take more time (both
for construction of the approximation, as well as for sampling) with
only a marginal improvement. For the maximum number of components,
$T=20$ was sufficient in the considered examples. All subsequent
applications in this paper rely on these default assumptions. When
these choices fail, \textit{i.e.}, the algorithm fails to identify
parts of the probability mass, one can increase $n$, or decrease
$\delta$ and $\eps$. Another strategy is to increase the number of
starting values at iteration 0, if complete modes might have been
missed at the beginning.

The algorithm was implemented in the \texttt{R} computing language
\cite{R}. To avoid floating point errors we work on the log-scale
whenever possible. When one needs to work on the original scale, for
example, in the quadratic programming step or in the optimization of
$\log(r(\bm x))$, one can use $\pi^*(\bm x)=\exp(\log(\pi(\bm
x))-\log(M_t))$, where $\log(M_t) = \underset{\bm X_t}{\max}
\log(\pi(\bm x))$. In this way excessively small values are avoided.

\section{Applications}
\label{sec:appl}

\renewcommand{\baselinestretch}{1.25}
\begin{figure}
\centerline{\includegraphics[width=1\textwidth]{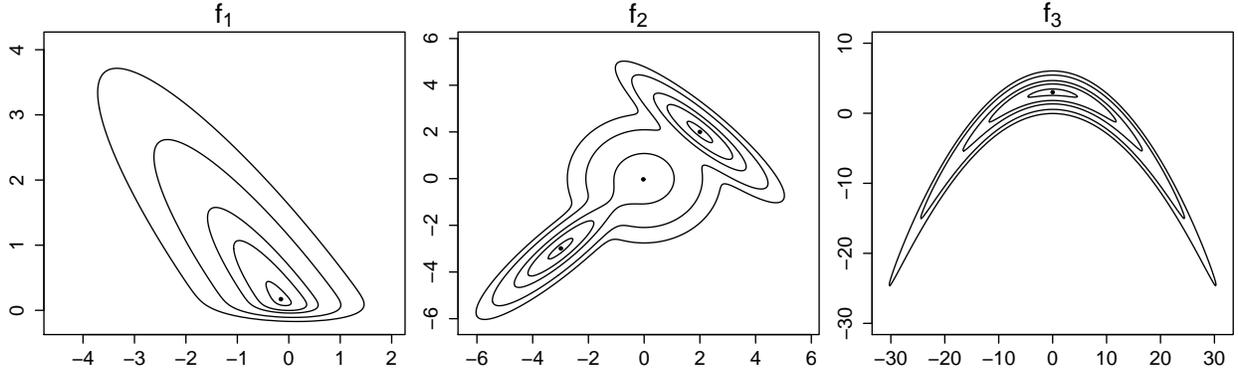}}
  \caption{Contour plot for the three test densities. For $f_3$ only
    the two first coordinates are plotted. Black dots ($\bullet$)
    denote local maxima and the contour lines are located at 0.9, 0.5,
    0.25, 0.05 and 0.01 of the height of the global maximum.}
  \label{fig:dens}
\end{figure}
\renewcommand{\baselinestretch}{2}

\subsection{Test Cases}
\label{sec:artif}

In this section the method is illustrated for three artificial, yet
realistic test cases. Common challenges for computational approaches
in Bayesian statistics are skew, non-linear and multimodal posterior
distributions. These often occur in applications beyond the standard
statistical models, for example, when the statistical model is not
from an exponential family with conjugate prior and linear predictors.

The first density $f_1$ is a bivariate skew $t-$distribution with 5
degrees of freedom, scale matrix $\bm \Omega =
\left( \begin{smallmatrix} 1 & -0.9 \\ -0.9 & 1 \end{smallmatrix}
\right)$, location vector $\bm \xi = (0,0)'$ and skewness vector $\bm
\alpha = (0, 15)'$ (see \citeasnoun{azza:capi:2003} for details on the
parametrization). This density (displayed in Figure \ref{fig:dens}
left) possesses an extreme non-elliptical skew shape. The second
density $f_2$ is a mixture of three bivariate normal distributions:
$f_2(\bm x)=0.34 \phi(\bm x, (0,0)', \bm S_1) + 0.33\phi(\bm x,
(-3,-3)', \bm S_2) + 0.33\phi(\bm x, (2,2)', \bm S_3)$, with $\bm S_1=
\left( \begin{smallmatrix} 1 & 0 \\ 0 & 1 \end{smallmatrix} \right) $,
$\bm S_2= \left( \begin{smallmatrix} 1 & 0.9 \\ 0.9 &
    1 \end{smallmatrix} \right) $ and $\bm S_3=
\left( \begin{smallmatrix} 1 & -0.9 \\ -0.9 & 1 \end{smallmatrix}
\right)$ \cite{gilk:robe:sahu:1998}. This density is multimodal and
has a complex local structure, see Figure \ref{fig:dens} (middle). The
third density is the ten dimensional non-linear banana shaped
distribution used in Wraith et al.~(2009)\nocite{wrai:kilb:bena:2009},
with density $f_3(\bm x) \propto \phi(t(\bm x), \bm m, \bm S)$, where
$t(\bm x) = (x_1,x_2+b(x_1^2-\sigma_1^2),x_3,\ldots,x_{10})$, $\bm
m=(0,\ldots,0)'$ and $\bm S = \mathrm{diag}(\sigma_1^2,1,\ldots,1)$,
$b=0.03$ and $\sigma^2_{1}=100$ were used, as in Wraith et
al.~(2009)\nocite{wrai:kilb:bena:2009}. Figure \ref{fig:dens} right,
displays the density of the first two components, when the other
coordinates are fixed at 0.

\begin{figure}
\centerline{\includegraphics[width=0.5\textwidth]{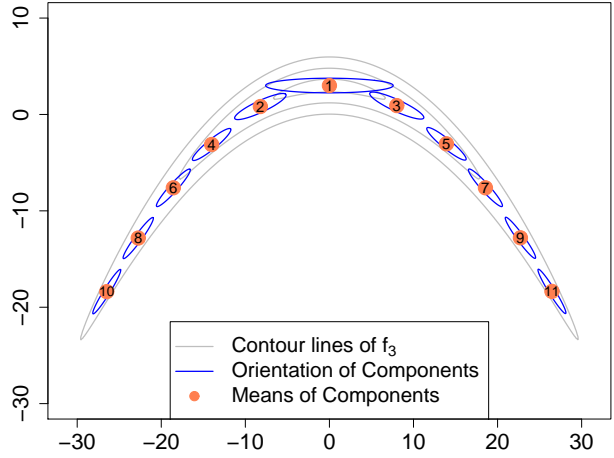}}
  \caption{Illustration of \textsf{iterLap} for the first two
    dimensions of $f_3$. The numbers in the dots refer to the order in
    which components were added.}
  \label{fig:illus}
\end{figure}

\textsf{iterLap} has been applied to these problems with the default
tuning parameters.  As starting values for the optimization in the
initial Laplace approximation in each case one vector consisting only
of zeros has been used. For $f_1$ \textsf{iterLap} selects 9
components and stops the iterative process, because the algorithm
cannot further change the approximation of the normalization
constant. For $f_2$ the algorithm stops the iterative process after 5
components, because the maximum error on the grid points is
achieved. For $f_3$ the iterative process is illustrated in some
detail in Figure \ref{fig:illus}, which displays the selected 11
components and the order in which they are selected. The algorithm
stops, because the normalization constant of the approximation does
not further change considerably. In Figure \ref{fig:illus} one can
also see that the orientation of the selected components' ellipses
fits the underlying local structure of the distribution quite well.

\renewcommand{\baselinestretch}{1.25}
\renewcommand{\baselinestretch}{2}

As distance measure to the true density the normalized effective
sample size ($\mathrm{NESS}$) has been calculated by an application of
importance sampling with the obtained mixture of normal distributions
as proposal. $\mathrm{NESS}$ lies in $(0,1]$, where larger values
correspond to a better fit and NESS is an estimate of a monotonic
transformation of the $\chi^2$-distance between proposal and true
distribution \cite{kong:liu:wong:1994}. It is defined as
$\mathrm{NESS}=1/(N\sum_{i=1}^N \tilde{\omega}_i^2),$ where $N$ is the
number of simulated values and $\tilde{\omega}_i$ the normalized
importance weights. The reported values of $\mathrm{NESS}$ in Table
\ref{tab:res3} are the average over 100 independent runs of importance
sampling each with sample size $N=10000$ (standard deviation given in
brackets). In addition the means and standard deviations of the
marginal distributions have been compared. For this purpose the
absolute distance between true and approximated mean and standard
deviation were used and divided by the true standard deviation of the
corresponding marginal distribution. Deviations from the true values
are thus measured in units of the true standard deviation.

\renewcommand{\baselinestretch}{1.25}
\begin{table}
  \centering
  \begin{tabular}[c]{ll|cccccc}
    \multicolumn{1}{c|} {Density}& Method &  $\mathrm{NESS}$ & Mean$_{x_1}$  &
    sd$_{x_1}$ & Mean$_{x_2}$  & sd$_{x_2}$  \\\hline\hline
    $f_1$ & Laplace   & 0.04 (0.04) & 0.72 & 0.49 & 0.88 & 0.59  \\
    & iter. Lapl.     & 0.65 (0.20) & 0.02 & 0.16 & 0.05 & 0.11 \\\hline
    $f_2$ & Laplace   & 0.02 (0.01) & 0.13 & 0.56 & 0.13 & 0.56  \\
    & iter. Lapl.     & 0.99 ($<$0.01)& $<$0.01 & $<$0.01 & $<$0.01 & $<$0.01  \\\hline
    $f_3$ &  Laplace  & 0.05 (0.04)  & $<$0.01 & $<$0.01 & 0.70 & 0.77  \\
    & iter. Lapl.     & 0.71 (0.04)  & $<$0.01 & 0.14 & 0.15 & 0.08 \\\hline
  \end{tabular}
  \caption{Normalized Effective Sample Size (NESS) and approximation
    error in the marginal mean and standard deviation relative to the
    true standard deviation for the Laplace
    approximation and the iterated Laplace approximation of $f_1$, $f_2$ and $f_3$.}
  \label{tab:res3}
\end{table}
\renewcommand{\baselinestretch}{2}

Table \ref{tab:res3} displays the results for \textsf{iterLap} and for
a standard Laplace approximation (which was used as the starting
approximation for \textsf{iterLap}).  One can conclude that the
iterated Laplace approximation has a substantially better performance
than the standard Laplace approximation particularly in terms of
$\mathrm{NESS}$, but also for the moments of the marginal
distributions.

The results regarding the ten dimensional $f_3$ are also quite
encouraging: Compared to the results of Wraith et al.~(2009, Figure
3), who use adaptive importance sampling based on
\citeasnoun{capp:douc:guil:2008}, one can observe that a similar
median effective sample size was obtained for 10 iterations of
adaptive importance sampling. The adaptive importance sampling,
however, has a larger variability and needs a total of 100000 function
evaluations in each case. Our approach needs a total of around 19000
function evaluations for building the approximation, including
evaluations for building the grid as well as evaluations needed for
the optimizer and calculating the Hessians. The number of function
evaluations is an important machine independent indicator on how fast
an algorithm runs, as the other computations needed by the algorithms
can usually be neglected, particularly if evaluation of the target
distribution is computationally expensive. To get an idea of the
actual computation time needed by our \texttt{R} implementation (which
does not exploit that parts of the code can be parallelized): Building
the global approximation for $f_3$ takes around 2.5 seconds (using a
Laptop computer with 1.86 Ghz and 2GB RAM).

\subsection{Nonlinear Regression}
\label{sec:ex}

\begin{figure}
\centerline{\includegraphics[width=0.6\textwidth]{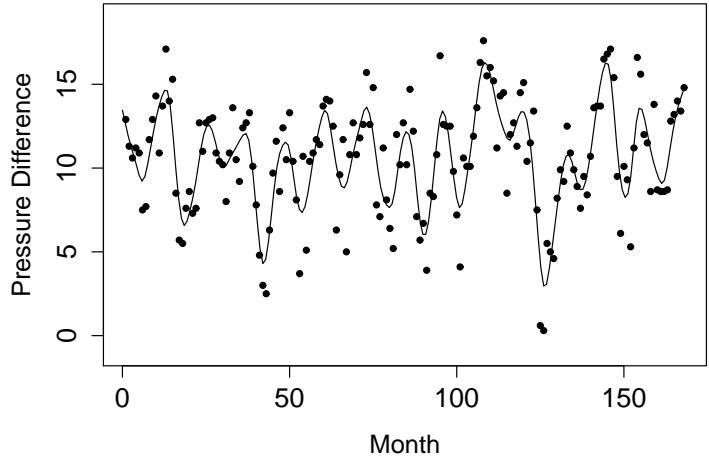}}
\caption{Pressure Difference between the Easter Islands and Darwin,
  Australia with smoothing spline fit (black line) to give an idea of
  the conditional mean function.}
  \label{fig:data}
\end{figure}

To illustrate \textsf{iterLap} on a real problem, data on monthly
averaged atmospheric pressure differences between the Easter Islands
and Darwin, Australia over 168 months are used (see Figure
\ref{fig:data}). These data are taken from the NIST website {\small
  www.itl.nist.gov/div898/strd/nls/data/enso.shtml}. The difference in
pressure is of meteorologic importance as it drives the trade winds in
the southern hemisphere, and the main purpose of the data analysis is
to infer the frequency of periodic cycles. The model for the data is
$y_i \sim N(\mu(i),\sigma^2)$, where
$\mu(i)=\alpha+\sum_{k=1}^3A_k\sin(2\pi i/\lambda_k)+B_{k}\cos(2\pi
i/\lambda_k)$ for $i=1,\ldots,168.$ For the conditionally linear
parameters $\alpha,A_k,B_k,\;k=1,2,3$ independent Cauchy a-priori
distributions with median 0 and scale 100 (for $\alpha$) and 10 (for
$A_k,B_k,\;k=1,2,3$) were used. For the positive parameter $\lambda_k$
independent uniform distributions on $[0,100]$ were employed and for
$\sigma$ a gamma distribution with parameters $0.1$ and $0.1$. The
transformation $\log(\sigma)$ has been used, to obtain a parameter that
lies in $\mathbb{R}$. All parameters will be summarized in the vector
$\bm \theta$. The likelihood surface is highly multimodal, but there
seems to be one dominant mode. As starting value for the optimization
the least squares estimate shown on the NIST website is used (note
however that here also $\lambda_1$ is treated as unknown, while in
their analysis this is fixed to 12).

\renewcommand{\baselinestretch}{1.25}
\renewcommand{\baselinestretch}{2}

To compare \textsf{iterLap} to other computational methods, four long
MCMC runs were produced (each started at the mode and of size 2500000
with thinning 10 after a burn-in of 10000) and the resulting 1000000
iterations are used as a gold-standard. These runs were obtained with
a multivariate random walk Metropolis algorithm implemented in the
\texttt{mcmc} package of \citeasnoun{geye:2010}, with proposal
variance matrix $2.38^2/11\tilde{\bm \Sigma},$ where $\tilde{\bm
  \Sigma}$ is the negative inverse Hessian at the mode of the
log-density.

Using the default tuning parameters, iterated Laplace approximation
selects 12 components and stops the iterative process, because the
normalization constant of the approximation does not further
improve. For this application the obtained distribution will be used
as a proposal distribution for importance sampling. As is common in
importance sampling, the Gaussian approximations obtained from
\textsf{iterLap} will be replaced by mixtures of t-distributions with
equal centering vector and scale matrix and 10 degrees of freedom,
because those possess heavier tails. To obtain an unweighted sample,
importance sampling resampling with residual resampling was used with
sample size 5000 (see \citeasnoun[Chapter 14]{robe:case:2004} for
details on residual resampling).

The \textsf{iterLap} procedure will be compared with two MCMC-based
approaches. First the componentwise adaptive random walk Metropolis
algorithm is used (see \citeasnoun{robe:rose:2009}). To improve the
componentwise updating, the algorithm is applied on the transformed
scale $\tilde{\bm \theta}=\tilde{\bm \Sigma}^{1/2}\bm \theta+
\tilde{\bm \mu}$, with $\tilde{\bm \mu}$ the posterior mode and
$\tilde{\bm \Sigma}^{1/2}$ the square root of the negative inverse
Hessian at the mode calculated from the eigen decomposition. This has
the advantage that the covariance matrix is approximately diagonal on
the transformed scale. For implementation the function
\texttt{adaptMetropGibbs} from the \texttt{spBayes} package
\cite{finl:bane:carl:2010} was used with target acceptance rate set to
0.44 and initial proposal standard deviations set to $2.38$. The
second MCMC approach is an optimally tuned multivariate random walk
Metropolis algorithm. The covariance matrix of the proposal was chosen
proportional to the covariance matrix obtained from the gold-standard
runs above, with the scaling selected as $2.38^2/11$. This is an
algorithm that cannot be used in practice, because the ``true''
covariance matrix is unavailable a-priori. Nevertheless, this
algorithm is of interest as it provides an upper bound on the
performance of the adaptive random walk Metropolis algorithm (as
described for example in \citeasnoun{robe:rose:2009}), which needs to
estimate the covariance matrix from its own iterations. In all cases
the starting value is chosen equal to the mode of the posterior
distribution.

\renewcommand{\baselinestretch}{1.25}
\begin{figure}
\centerline{\includegraphics[width=1\textwidth]{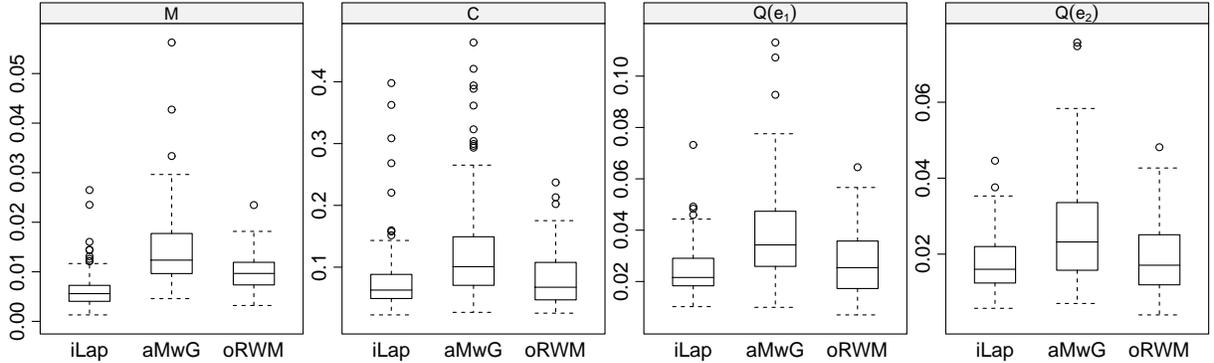}}
\caption{Simulation results for the three methods over 100
  repetitions. \textsf{iLap} $\hat{=}$ importance sampling with
  \textsf{iterLap}, \textsf{aMwG.} $\hat{=}$ adaptive Metropolis
  within Gibbs, \textsf{oRWM} $\hat{=}$ optimal random walk Metropolis
  with true covariance matrix.}
  \label{fig:resReal}
\end{figure}
\renewcommand{\baselinestretch}{2}

Four performance measures are used to compare the
methodologies. First, the estimates of the posterior means are
compared by calculating the Mahalanobis distance $M(\hat{\bm
  \theta},\bm \theta_{true})=\sqrt{(\hat{\bm \theta}-\bm
  \theta_{true})'\bm S_{true}^{-1}(\hat{\bm \theta}-\bm
  \theta_{true})}$, between the empirical average $\hat{\bm \theta}$
from the simulation output and the values $\bm \theta_{true}$ and $\bm
S_{true}$ obtained from the gold-standard runs. To compare the
estimates of the parameter covariance matrix, the spectral norm of the
difference between the empirical and true covariance matrices
$C(\hat{\bm S},\bm S_{true})=\sqrt{\lambda_{max}(\bm D'\bm D)}$ was
calculated, where $\bm D=\hat{\bm S}-\bm S_{true}$ and
$\lambda_{max}(.)$ is the function that returns the largest
eigenvalue.  To obtain measures of multivariate goodness of fit, the
quantiles
$q^{(j)}_{0.05},q^{(j)}_{0.1},q^{(j)}_{0.15},\ldots,q^{(j)}_{0.95}$
along the first two eigenvectors $e_j,\;j=1,2$ of the true correlation
matrix were calculated from the gold-standard runs and then compared
to the quantiles obtained from the simulations
$Q(e_j)=\frac{1}{19}\underset{i\in \{0.05,0.1,\ldots,0.95\}}{\sum}
|\hat{q}^{(j)}_{i}-q_i^{(j)}|,$ for $j=1,2$.

All three algorithms were applied to the problem and repeated 100
times. The \textsf{iterLap} procedure requires around 28000 function
evaluations values in total (both for building the approximation and
the 5000 evaluations for importance sampling). For the MCMC based
algorithms twice as many function evaluations namely a total of 60000
iterations were allowed, with 10000 burn-in and thinning rate 10, so
that in summary also 5000 iterations are obtained. From the results
displayed in Figure \ref{fig:resReal} one can conclude that importance
sampling with \textsf{iterLap} works very well compared to the other
approaches in all performance measures (particularly for the posterior
moments), although it uses fewer function evaluations.  In addition it
is easy to obtain a reliable estimate of the normalizing constant via
importance sampling, while this is more complicated to obtain reliably
from MCMC output.

Returning to the original aim of the data analysis, the resulting
posterior means for the frequency parameters
$(\lambda_1,\lambda_2,\lambda_3)$ are given by $(11.9, 44.1, 26.8)'$
months with standard deviations $(0.04, 1.1, 0.36)'$. These
frequencies can be traced to the yearly cycle ($\lambda_1$), El
Ni\~{n}o ($\lambda_2$) and the Southern Oscillation ($\lambda_3$), see
the NIST website for details.

\section{Discussion}
\label{sec:disc}

Compared to traditional function approximation or regression function
estimation, globally approximating a positive integrable function
$\pi(\bm x)$ proportional to a probability density is a considerably
more difficult problem. The main complication is that it is a-priori
unclear \textit{where} to approximate $\pi(\bm x)$, \textit{i.e.},
where most of the probability mass of $\pi(\bm x)$ is located. In this
article the iterated Laplace approximation has been introduced to
solve this twofold problem of identification of regions with relevant
probability mass, and approximation of $\pi(\bm x)$ in these
regions. The methodology starts with a simple Laplace approximation,
and then iteratively applies Laplace approximation to the residual
between truth and current approximation, until a stopping criterion is
satisfied. By optimizing the residual in each step of the procedure,
the algorithm identifies regions with relevant probability mass, where
the current approximation fits poorly and an improvement is
needed. Once a mode and the local curvature is determined, the new
component is added to the approximation, and the coefficients are
determined by minimizing the $L_2$ distance between truth and
approximation on a grid.

In this paper the methodology has been evaluated in three test cases
and one real example with positive results. In the case of the ten
dimensional banana shaped example, the approach obtained similar
results as adaptive importance sampling based on a mixture of t
distributions with improved computational efficiency in terms of
function evaluations. Further, in the real data example, where two
state of the art MCMC algorithms were applied, the iterated Laplace
approximation showed a very competitive performance with a smaller
number of function evaluations.

As for all analytical approximations, it is difficult to assess the
quality of the obtained iterated Laplace approximation in a concrete
modelling situation. Hence, its main value is to use it as a proposal
distribution for Monte Carlo techniques. These techniques allow to
assess the quality of the approximation and correct for deficiencies
of fit by rejecting or weighting samples. While the focus in this
paper has been on importance sampling techniques (where the effective
sample size can be used to assess the quality of the approximation),
one can, of course, also use the approximation in the context of MCMC
techniques that employ a global proposal distribution, such as the
independence Metropolis-Hastings algorithm or the rejection sampling
Metropolis-Hastings algorithm \cite[ch. 2.3]{tier:1994}, in these
cases reliable MCMC standard errors can be calculated to evaluate the
quality of the simulation (see \citeasnoun{fleg:hara:jone:2008}).

A primary application of the methodology might be non-linear models as
applied in diverse fields, for example early phase clinical trials or
cosmology. In these models the posterior distribution can be skew and
multimodal, and one usually cannot design Gibbs moves to directly
sample from the full conditional distributions. Nevertheless, the
algorithm was also tested with success on a variety of other
applications, for example dose-response estimation, Gaussian process
regression and simple hierarchical models.

In higher dimensional problems it gets difficult to build a global
approximation of non-trivial posterior distributions, and the proposed
methodology is no exception: In these cases local MCMC moves often
become more efficient, although the iterated Laplace approximation
typically still provides an improvement over the standard Laplace
approximation in terms of building a global approximation of the
posterior and approximating the normalization constant. A
computational concern with regard to the methodology is the need for
numerically calculating the Hessian matrices. Depending on the
problem, this might become unstable (for example when the objective
function is flat in the neighborhood of the mode, or the mode lies on
a ridge) and, in larger dimensions, computationally expensive. A
partial solution, as suggested by a referee, is to use a structured
form of the covariance matrix (\textit{e.g.}, a diagonal matrix). This
reduces the computational burden in high dimensional cases and
stabilizes computations. The downside would be that a worse fit is
obtained by the added components and it is likely that more mixture
components are required to obtain an adequate approximation. Another
challenge for the \textsf{iterLap} methodology are situations when the
target density contains a large number of strongly separated modes. A
partial solution in these cases is to use more widely dispersed
starting values for the starting approximation (at iteration
0). Alternatively, one could also consider to use tempered version of
the residual function to avoid getting trapped in minor local modes.

\textbf{Acknowledgements}\\
The author would like thank Frank Bretz and David Ohlssen for
proofreading the manuscript, and the Editor, Associate Editor and two
referees for helpful comments and suggestions that improved the
presentation of this article.

\bibliographystyle{diss}\bibliography{bibl}

%<!-- Local IspellDict: english -->
%<!-- Local IspellPersDict: ~/emacs/.ispell-english -->
\end{document}